\input epsf
\newfam\scrfam
\batchmode\font\tenscr=rsfs10 \errorstopmode
\ifx\tenscr\nullfont
        \message{rsfs script font not available. Replacing with calligraphic.}
        \def\scr{\cal}
\else   
        \font\sevenscr=rsfs7
        \font\fivescr=rsfs5
        \skewchar\tenscr='177 \skewchar\sevenscr='177 \skewchar\fivescr='177
        \textfont\scrfam=\tenscr \scriptfont\scrfam=\sevenscr
        \scriptscriptfont\scrfam=\fivescr
        \def\scr{\fam\scrfam}
        \def\cal{\scr}
\fi
\catcode`\@=11
\newfam\frakfam
\batchmode\font\tenfrak=eufm10 \errorstopmode
\ifx\tenfrak\nullfont
        \message{eufm font not available. Replacing with italic.}
        \def\frak{\it}
\else
	
	\font\sevenfrak=eufm7 \font\fivefrak=eufm5
	\textfont\frakfam=\tenfrak
	\scriptfont\frakfam=\sevenfrak \scriptscriptfont\frakfam=\fivefrak
	\def\frak{\fam\frakfam}
\fi
\catcode`\@=\active
\newfam\msbfam
\batchmode\font\twelvemsb=msbm10 scaled\magstep1 \errorstopmode
\ifx\twelvemsb\nullfont\def\Bbb{\bf}

	\message{Blackboard bold not available. Replacing with boldface.}
\else   \catcode`\@=11
        \font\tenmsb=msbm10 \font\sevenmsb=msbm7 \font\fivemsb=msbm5
        \textfont\msbfam=\tenmsb
        \scriptfont\msbfam=\sevenmsb \scriptscriptfont\msbfam=\fivemsb
        \def\Bbb{\relax\expandafter\Bbb@}
        \def\Bbb@#1{{\Bbb@@{#1}}}
        \def\Bbb@@#1{\fam\msbfam\relax#1}
        \catcode`\@=\active

\fi
        \font\eightrm=cmr8              \def\xrm{\eightrm}
        \font\eightbf=cmbx8             \def\xbf{\eightbf}
        \font\eightit=cmti10 at 8pt     \def\xit{\eightit}
                     
        \font\eightcp=cmcsc8
        \font\eighti=cmmi8              \def\xold{\eighti}
        \font\eightib=cmmib8             \def\xbold{\eightib}
        \font\teni=cmmi10               \def\old{\teni}
        \font\tencp=cmcsc10
        \font\tentt=cmtt10
        
        \font\twelvecp=cmcsc10 scaled\magstep1

	 at10pt	
	\font\twelvehelvbold=phvb at12pt
	 at14pt
	\font\sixteenhelvbold=phvb at16pt

\def\noblackbox{\overfullrule=0pt}
\noblackbox

\newtoks\headtext
\headline={\ifnum\pageno=1\hfill\else
	\ifodd\pageno{\eightcp\the\headtext}{ }\dotfill{ }{\old\folio}
	\else{\old\folio}{ }\dotfill{ }{\eightcp\the\headtext}\fi
	\fi}
\def\makeheadline{\vbox to 0pt{\vss\noindent\the\headline\break
\hbox to\hsize{\hfill}}
        \vskip2\baselineskip}
\newcount\infootnote
\infootnote=0
\def\foot#1#2{\infootnote=1
\footnote{${}^{#1}$}{\vtop{\baselineskip=.75\baselineskip
\advance\hsize by -\parindent\noindent{\xrm #2}}}\infootnote=0$\,$}
\newcount\refcount
\refcount=1
\newwrite\refwrite
\def\oldsize{\ifnum\infootnote=1\xold\else\old\fi}
\def\ref#1#2{
	\def#1{{{\oldsize\the\refcount}}\ifnum\the\refcount=1\immediate\openout\refwrite=\jobname.refs\fi\immediate\write\refwrite{\item{[{\xold\the\refcount}]} 
	#2\hfill\par\vskip-2pt}\xdef#1{{\noexpand\oldsize\the\refcount}}\global\advance\refcount by 1}
	}
\def\refout{\catcode`\@=11
        \xrm\immediate\closeout\refwrite
        \vskip2\baselineskip
        {\noindent\twelvecp References}\hfill\vskip\baselineskip
        \baselineskip=.75\baselineskip
        \input\jobname.refs
        \baselineskip=4\baselineskip \divide\baselineskip by 3
        \catcode`\@=\active\rm}

\def\hepth#1{\href{http://xxx.lanl.gov/abs/hep-th/#1}{hep-th/{\xold#1}}}
\def\jhep#1#2#3#4{\href{http://jhep.sissa.it/stdsearch?paper=#2\%28#3\%29#4}{J. High Energy Phys. {\xbold #1#2} ({\xold#3}) {\xold#4}}}
\def\AP#1#2#3{Ann. Phys. {\xbold#1} ({\xold#2}) {\xold#3}}
\def\CMP#1#2#3{Commun. Math. Phys. {\xbold#1} ({\xold#2}) {\xold#3}}
\def\CQG#1#2#3{Class. Quantum Grav. {\xbold#1} ({\xold#2}) {\xold#3}}

\def\JHEP{\jhep}

\def\MPLA#1#2#3{Mod. Phys. Lett. {\xbf A}{\xbold#1} ({\xold#2}) {\xold#3}}
\def\NPB#1#2#3{Nucl. Phys. {\xbf B}{\xbold#1} ({\xold#2}) {\xold#3}}

\def\PLB#1#2#3{Phys. Lett. {\xbf B}{\xbold#1} ({\xold#2}) {\xold#3}}

\def\PRD#1#2#3{Phys. Rev. {\xbf D}{\xbold#1} ({\xold#2}) {\xold#3}}

\newcount\sectioncount
\sectioncount=0
\def\section#1#2{\global\eqcount=0
	\global\subsectioncount=0
        \global\advance\sectioncount by 1
	\ifnum\sectioncount>1
	        \vskip2\baselineskip
	\fi
	\noindent
        \line{\twelvecp\the\sectioncount. #2\hfill}
		\vskip.8\baselineskip\noindent
        \xdef#1{{\old\the\sectioncount}}}
\newcount\subsectioncount
\def\subsection#1#2{\global\advance\subsectioncount by 1
	\vskip.8\baselineskip\noindent
	\line{\tencp\the\sectioncount.\the\subsectioncount. #2\hfill}
	\vskip.5\baselineskip\noindent
	\xdef#1{{\old\the\sectioncount}.{\old\the\subsectioncount}}}
\newcount\appendixcount
\appendixcount=0
\def\appendix#1{\global\eqcount=0
        \global\advance\appendixcount by 1
        \vskip2\baselineskip\noindent
        \ifnum\the\appendixcount=1
        \hbox{\twelvecp Appendix A: #1\hfill}\vskip\baselineskip\noindent\fi
    \ifnum\the\appendixcount=2
        \hbox{\twelvecp Appendix B: #1\hfill}\vskip\baselineskip\noindent\fi
    \ifnum\the\appendixcount=3
        \hbox{\twelvecp Appendix C: #1\hfill}\vskip\baselineskip\noindent\fi}
\def\acknowledgements{\vskip2\baselineskip\noindent
        \underbar{\it Acknowledgements:}\ }
\newcount\eqcount
\eqcount=0
\def\Eqn#1{\global\advance\eqcount by 1
\ifnum\the\sectioncount=0
	\xdef#1{{\oldsize\the\eqcount}}
	\eqno({\oldstyle\the\eqcount})
\else
        \ifnum\the\appendixcount=0
	        \xdef#1{{\old\the\sectioncount}.{\old\the\eqcount}}
                \eqno({\oldstyle\the\sectioncount}.{\oldstyle\the\eqcount})\fi
        \ifnum\the\appendixcount=1
	        \xdef#1{{\oldstyle A}.{\old\the\eqcount}}
                \eqno({\oldstyle A}.{\oldstyle\the\eqcount})\fi
        \ifnum\the\appendixcount=2
	        \xdef#1{{\oldstyle B}.{\old\the\eqcount}}
                \eqno({\oldstyle B}.{\oldstyle\the\eqcount})\fi
        \ifnum\the\appendixcount=3
	        \xdef#1{{\oldstyle C}.{\old\the\eqcount}}
                \eqno({\oldstyle C}.{\oldstyle\the\eqcount})\fi
\fi}
\def\Fqn#1#2{\global\advance\eqcount by 1
\ifnum\the\sectioncount=0
	\xdef#1{{\old\the\eqcount}}
	\xdef#2{{\xold\the\eqcount}}
	\eqno({\oldstyle\the\eqcount})
\else
	\xdef#1{{\old\the\sectioncount}.{\old\the\eqcount}}
	\xdef#2{{\xold\the\sectioncount}.{\xold\the\eqcount}}
	\eqno({\oldstyle\the\sectioncount}.{\oldstyle\the\eqcount})
\fi}
\def\eqn{\global\advance\eqcount by 1
\ifnum\the\sectioncount=0
	\eqno({\oldstyle\the\eqcount})
\else
        \ifnum\the\appendixcount=0
                \eqno({\oldstyle\the\sectioncount}.{\oldstyle\the\eqcount})\fi
        \ifnum\the\appendixcount=1
                \eqno({\oldstyle A}.{\oldstyle\the\eqcount})\fi
        \ifnum\the\appendixcount=2
                \eqno({\oldstyle B}.{\oldstyle\the\eqcount})\fi
        \ifnum\the\appendixcount=3
                \eqno({\oldstyle C}.{\oldstyle\the\eqcount})\fi
\fi}
\def\multi{\global\advance\eqcount by 1}
\def\multieq#1#2{\xdef#1{{\old\the\eqcount#2}}
        \eqno{({\oldstyle\the\eqcount#2})}}
\newtoks\url
\def\Href#1#2{\catcode`\#=12\url={#1}\catcode`\#=\active#2}
\def\href#1#2{{#2}}
\def\hhref#1{{#1}}
\parskip=3.5pt plus .3pt minus .3pt
\baselineskip=14pt plus .1pt minus .05pt
\lineskip=.5pt plus .05pt minus .05pt
\lineskiplimit=.5pt
\abovedisplayskip=18pt plus 4pt minus 2pt
\belowdisplayskip=\abovedisplayskip
\hsize=14cm
\vsize=19cm
\voffset=1.8cm
\frenchspacing
\footline={}
\raggedbottom

\def\ss{\scriptstyle}
\def\sss{\scriptscriptstyle}
\def\*{\partial}
\def\punkt{\,\,.}
\def\komma{\,\,,}

\def\={\!=\!}
\def\small#1{{\hbox{$#1$}}}
\def\half{\small{1\over2}}
\def\fraction#1{\small{1\over#1}}
\def\fr{\fraction}
\def\Fraction#1#2{\small{#1\over#2}}
\def\Fr{\Fraction}
\def\tr{\hbox{\rm tr}}
\def\eg{{\tenit e.g.}}

\def\ie{{\tenit i.e.}}

\def\nlni{\hfill\break}

\def\a{\alpha}

\def\d{\delta}
\def\e{\varepsilon}

\def\O{\Omega}

\def\Z{{\Bbb Z}}

\def\Ham{{\cal H}}

\def\Tr{\hbox{Tr}\,}

%
%

\def\s{\sigma}
\def\r{\varrho}
\def\rr{{\r\over2\pi}}
\def\e{\varepsilon}

\def\A{{\cal A}}
\def\B{{\cal B}}

\def\T{{\cal T}}
\def\O{{\cal O}}

\def\d{\partial}

\def\Z{{\Bbb Z}}

\def\II{\hbox{I\hskip-0.6pt I}}

\def\mod{\,\,\hbox{mod}\,\,}

%
%
%

\headtext={Martin Cederwall: ``Open and Winding Membranes...''}

\line{
\epsfysize=1.7cm
\epsffile{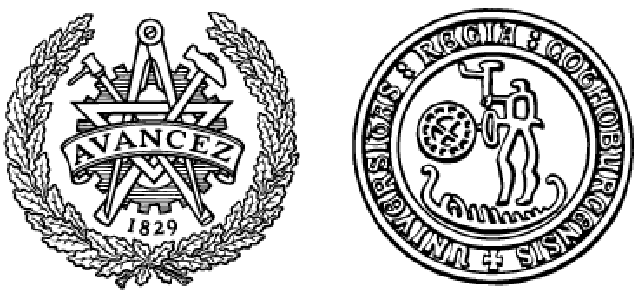}
\hfill}
\vskip-1.7cm
\line{\hfill G\"oteborg ITP preprint}
\line{\hfill hep-th/0210152}
\line{\hfill October, {\old2002}}
\line{\hrulefill}

\vfill

\centerline{\sixteenhelvbold Open and Winding Membranes, 
Affine Matrix Theory} 
\vskip.3cm
\centerline{\sixteenhelvbold and Matrix String Theory}

\vskip1.6cm

\centerline{\twelvehelvbold Martin Cederwall}

\vskip.8cm

\centerline{\it Department of Theoretical Physics}
\centerline{\it G\"oteborg University and Chalmers University of Technology }
\centerline{\it S-412 96 G\"oteborg, Sweden}

\vskip1.6cm

{\narrower\noindent 
\underbar{Abstract:} We examine the structure of winding toroidal and
open cylindrical membranes, especially in cases where 
they are stretched between boundaries. 
Non-zero winding or
stretching means that there are linear terms in the mode expansion of 
the coordinates obeying Dirichlet boundary conditions. A linear term 
acts as an outer derivation on the subalgebra of volume-preserving 
diffeomorphisms generated by single-valued functions, and obstructs 
the truncation to matrix theory obtained via non-commutativity with 
rational parameter. As long as only one of the two membrane 
directions is stretched, the possible consistent truncation is to 
coordinates taking values in representations of an affine algebra.
We show that this consistent truncation of the supermembrane gives
a precise microscopic derivation of matrix string theory with the 
representation content appropriate for the physical situation.
The matrix superstring theory describing parallel M{\old5}-branes is derived.
We comment on the possible applications of the construction to membrane 
quantisation in certain M-theory backgrounds.
\smallskip}
\vfill

\line{\hrulefill}
\catcode`\@=11
\line{\tentt martin.cederwall@fy.chalmers.se\hfill}
\catcode`\@=\active
\line{\tentt \hhref{http://fy.chalmers.se/tp/cederwall/}\hfill}

\eject

\ref\deWitHoppeNicolai{B. de Wit, J. Hoppe and H. Nicolai,
	``On the quantum mechanics of supermembranes'',
	\NPB{305}{1988}{545}.}

\ref\deWitLuscherNicolai{B. de Wit, M L\"uscher and H. Nicolai,
	``The supermembrane is unstable'',
	\NPB{320}{1989}{135}.}

\ref\deWitMarquardNicolai{B. de Wit, U. Marquard and H. Nicolai,
	``Area preserving diffeomorphisms and supermembrane 
	Lorentz invariance'', \CMP{128}{1990}{39}.}

\ref\CederwallOpenMembrane{M. Cederwall, 
	``Boundaries of {\xold11}-dimensional membranes'', 
	\MPLA{12}{1997}{2641} [\hepth{9704161}].}

\ref\HoravaWitten{P. Ho\v rava and E. Witten,
	``Heterotic and type I string dynamics from eleven-dimensions'',
	\NPB{460}{1996}{506} [\hepth{9510209}];
	``Eleven-dimensional supergravity on a manifold with boundary'',
	\NPB{475}{1996}{94} [\hepth{9603142}].}

\ref\FairlieZachos{D.B. Fairlie and C.K. Zachos, 
	``Infinite dimensional algebras, sine brackets and 
	{\xit su}($\ss\infty$)'', \PLB{224}{1989}{101}.}

\ref\KimRey{N. Kim and S.-J. Rey,
	``M(atrix) theory on an orbifold and twisted membrane'',
	\NPB{504}{1997}{189} [\hepth{9701139}].}

\ref\EzawaMatsuoMurakami{K. Ezawa, Y. Matsuo and K. Murakami,
	``Matrix regularization of open supermembrane: 
	towards M theory five-brane via open supermembrane'',
	\PRD{57}{1998}{5118} [\hepth{9707200}].}

\ref\BeckerBecker{K. Becker and M. Becker, ``Boundaries in M theory'',
	\NPB{472}{1996}{221} [\hepth{9602071}].}

\ref\Hoppe{J. Hoppe, ``Zero energy states in supersymmetric matrix models'',
	\CQG{17}{2000}{1101}, and references therein.}

\ref\MatrixReviews{M.J. Duff, ``Supermembranes'', \hepth{9611203};
	\nlni H. Nicolai and R. Helling, 
		``Supermembranes and (M)atrix theory'', \hepth{9809103};
	\nlni B. de Wit, ``Supermembranes and super matrix models'', 
		\hepth{9902051}; 
	\nlni W. Taylor, ``(M)atrix theory: Matrix quantum mechanics 
		as fundamental theory'', \hepth{0101126};
	\nlni T. Banks, ``TASI lectures on matrix theory'', \hepth{9911068}.}

\ref\deWitPeetersPlefka{B. de Wit, K. Peeters and J. Plefka,
	``Supermembranes with winding'', 
	\PLB{409}{1997}{117} [\hepth{9705225}]}

\ref\BFSS{T. Banks, W. Fischler, S.H. Shenker and L. Susskind,
	``M theory as a matrix model: a conjecture'', 
	\PRD{55}{1997}{5112} [\hepth{9610043}].}

\ref\SethiStern{S. Sethi and M. Stern, ``D-brane bound states redux'',
		\CMP{194}{1998}{675} [\hepth{9705046}].}

\ref\KacSmilga{V.G. Kac and A.V. Smilga, 
		``Normalized vacuum states in N=4 supersymmetric 
		Yang--Mills quantum mechanics with any gauge group'',
		\NPB{571}{2000}{515} [\hepth{9908096}].}

\ref\BST{E. Bergshoeff, E. Sezgin and P.K. Townsend,
	``Supermembranes and eleven-dimensional supergravity'',
	\PLB{189}{1987}{75}; 
	``Properties of the eleven-dimensional super membrane theory'',
	\AP{185}{1988}{330}.}

\ref\BSTT{E. Bergshoeff, E. Sezgin, Y. Tanii and P.K. Townsend,
	``Super p-branes as gauge theories of 
	volume preserving diffeomorphisms'',
	\AP{199}{1990}{340}.}

\ref\ChuSezgin{C.-S. Chu and E. Sezgin, 
	``M five-brane from the open supermembrane'',
	\JHEP{97}{12}{1997}{001} [\hepth{9710223}].}

\ref\FuchsSchweigert{J. Fuchs and C. Schweigert, 
	``Symmetries, Lie algebras and representations'',
	Cambridge University Press, 1997.}

\ref\DVV{R. Dijkgraaf, E. Verlinde and H. Verlinde,
	``Matrix string theory'', \NPB{500}{1997}{43} [\hepth{9703030}].} 

\ref\SekinoYoneya{Y. Sekino and T. Yoneya, 
	``From supermembrane to matrix string'',
	\NPB{619}{2001}{22} [\hepth{0108176}].}

\ref\Minic{Dj. Mini\'c, ``M-theory and deformation quantization'',
	\hepth{9909022}.}	

\ref\BraxMourad{Ph. Brax and J. Mourad, 
	``Open supermembranes in eleven-dimensions''
	\PLB{408}{1997}{142} [\hepth{9704165}];
	``Open supermembranes coupled to M theory five-branes'',
	\PLB{416}{1998}{295} [\hepth{9707246}].}

\ref\Henningson{A. Gustavsson and M. Henningson, 
	``A short representation of the six-dimensional (2,0) algebra'',
	\JHEP{01}{06}{2001}{054} [\hepth{0104172}].}

\ref\BermanEtAl{D.S. Berman, M. Cederwall, U. Gran, H. Larsson,
      M. Nielsen, B.E.W. Nilsson and P. Sundell
	``Deformation independent open brane metrics and generalized
      theta parameters'',
	\jhep{02}{02}{2002}{012} [\hepth{0109107}].}

\ref\Schaar{J.P. van der Schaar, ``The reduced open membrane metric'',
	\jhep{01}{08}{2001}{048} [\hepth{0106046}].}

\section\intro{Introduction}Supermembrane [\BST] theory [\deWitHoppeNicolai]
is a very promising
candidate for a microscopic description of M-theory. Although it is not
background invariant, it gives a completely new picture of the nature of
space and time at small scales, together with a description of 
quantum-mechanical states that goes beyond local quantum field theory.
These features are most clear in the matrix [\BFSS] truncation 
[\deWitHoppeNicolai] of the membrane.
It is widely appreciated that first-quantised supermembrane theory 
through its continuous spectrum [\deWitLuscherNicolai] is capable of
describing an entire (``multi-particle'') Fock space. 
For reviews on the subject of membranes and matrices, see ref. 
[\MatrixReviews].
Due to the immense
technical difficulties associated with actual calculations in the theory, 
which is non-linear and inherently non-perturbative, few quantitative
features are known in addition to the general picture, which is supported 
by many qualitative arguments. The maybe most important one is the proof
that $su(N)$ matrix theory has a unique supersymmetric ground state
[\SethiStern,\KacSmilga], which gives
the relation to the massless degrees of freedom of $D=11$ supergravity.

Many situations in M-theory backgrounds involve membranes that are not closed.
Supermembranes may end on ``defects'', \ie, {\old5}-branes and {\old9}-branes
[\HoravaWitten,\CederwallOpenMembrane,\BraxMourad,\ChuSezgin,\BeckerBecker].
It is urgent to have some mathematical formulation of these situations
in order to understand the microscopic properties of physics in such
backgrounds. One old enigma is the nature of the theory on multiple
{\old5}-branes, which we hope to be able to address 
using the present techniques.
Matrix regularisation of winding and stretched membranes has been attempted
before [\deWitPeetersPlefka], but not entirely 
successfully. The matrix regularisation of open
unstretched supermembranes has been considered in ref. [\EzawaMatsuoMurakami].
The aim of this paper is to demonstrate
why truncation to finite matrices fails, and what the consistent truncation
instead is. It turns out to contain an affine algebra. A similar construction,
based on discretisation, rather than non-commutativity, and thus more
{\it ad hoc} postulated to match the structure of matrix string theory,
has been considered in ref. [\SekinoYoneya] for a winding membrane.
The present paper provides a natural microscopic explanation of that
construction.

We start by a careful examination of the boundary conditions for open 
membranes. Due to the non-linearity of membrane dynamics, this is a
non-trivial issue. In section {\old2}, which can be skipped by any reader
feeling confident with starting from imposed linear Dirichlet and Neumann
conditions, we show how these are obtained using a partial gauge fixing.
In section {\old3} we review
the consistent truncation of membranes to matrices through introduction
of a rational non-commutativity parameter, and generalise the procedure in
presence of winding or stretching to truncation to elements in an
affine algebra. Section {\old4} is devoted to the resulting matrix theory,
which is shown to be equivalent to matrix string theory with a certain
matter content, and in section {\old5} everything is extended to the
supersymmetric membrane. We investigate in particular the
half-supersymmetric case of open membranes ending on
M{\old5}-branes. 
Finally, section {\old6} discusses some 
possible applications, open questions and directions of future investigations.

\section\BoundaryConditions{Analysis of boundary conditions}We start
from the action for a bosonic membrane in flat $D$-dimensional 
Minkowski space with 
vanishing {\old3}-form tensor field,
$$
S=-T\int d^3\xi\sqrt{-g}\komma\eqn
$$
where $g$ is the the determinant of the induced metric 
$g_{IJ}=\d_IX^a\d_JX^b\eta_{ab}$.
The momenta conjugate to $X^a$ are
$$
P_a={\delta S\over\delta\dot X^a}=-T\sqrt{-g}(g^{00}\dot 
X_a+g^{0i}\d_iX_a)\komma\eqn
$$
where $i,j,\ldots=1,2$ are spacelike indices for the membrane and $g$ 
with upper indices denotes the inverse of $g_{IJ}$. The constraints 
resulting are the ones corresponding to space reparametrisations,
$$
\T_{i}\equiv P_a\d_iX^a\approx0\komma\eqn
$$
and the hamiltonian constraint
$$
\Ham\equiv\fr{2T}P^2+\Fr T2\tilde g\approx0\komma\eqn
$$
$\tilde g_{ij}=\d_iX^a\d_jX^b\eta_{ab}$ being the induced metric 
on the membrane space-sheet, 
and $\tilde g$ its determinant. It is conveniently rewritten in terms 
of the ``Poisson bracket'' on the space-sheet\foot\star{In this paper
we will specialise to toroidal and cylindrical membrane topology---in general
cases the Poisson bracket should include a scalar density in order to
map scalar functions to scalar functions.}, 
$$
\{A,B\}=\e^{ij}\d_iA\d_jB\komma\Eqn\PoissonBracket
$$
as $\tilde g=\half\{X_a,X_b\}\{X^a,X^b\}$, so that the hamiltonian 
constraint takes the form
$$
\Ham=\fr{2T}P^2+\Fr T2\{X_a,X_b\}\{X^a,X^b\}\approx0\punkt\eqn
$$

Consider an open membrane whose end-strings are 
confined to some (flat) hypersurfaces. We will soon choose the 
topology of a finite cylinder for the membrane, but for the moment 
the analysis applies locally at each boundary. For simplicity, we restrict our 
attention to parallel stationary hypersurfaces of space dimension $p$, 
and choose the coordinates of the 
embedding space so that the hypersurfaces are coordinate 
surfaces specified by $X^{a''}=X_{(i)}^{a''}$, $a''=p+1,\ldots,D-1$ 
for the boundary component $(i)$. We also choose local coordinates 
$\xi^i_{(i)}=(\s_{(i)},\r_{(i)})$ on 
the membrane space-sheet so that boundary $(i)$ is $\r_{(i)}=0$. 
Dropping the index $i$ and looking at one boundary component at the 
time, we thus have the $D-p-1$ Dirichlet boundary conditions 
$$
X^{a''}(\s,0)=x^{a''}\komma\quad 
a''=p+1,\ldots,D-1\punkt\Eqn\FirstDirichlet
$$
The boundary conditions for the remaining components may be obtained 
as momentum conservation in the longitudinal directions. Demanding 
that $p_{a'}=\int d^2\xi P_{a'}$, $a'=0,\ldots,p$, is conserved, \ie, that 
$\{H,p_{a'}\}_{PB}=0$,
where $H=\int d^2\xi\Ham$, so that there is no 
flow of momentum across the 
boundary, gives the Neumann-like conditions
$$
\eqalign{
&\d_1 X^{b'}\{X_{b'},X_{a'}\}(\s,0)\cr
&=\bigl[(\d_1 X)^2\d_2 X_{a'}-(\d_1 X\cdot\d_2 X)\d_1 X_{a'}\bigr](\s,0)=0
\komma\quad a'=0,\ldots,p\punkt\cr}\Eqn\FirstQuasiNeumann
$$

The boundary conditions (\FirstDirichlet) and (\FirstQuasiNeumann) are 
of course to be treated as constraints, and as such, one must assure that 
their time evolution is consistent. Taking the Poisson brackets with 
$H$ will yield new, secondary, constraints, and 
so on. This is a non-trivial procedure, since the membrane dynamics is 
non-linear.
First, we note that the non-linear boundary conditions on $X^{a'}$ 
state that the part of $\d_2X^{a'}$ orthogonal to $\d_1X^{a'}$ 
vanishes. Such conditions become in practice intractable, and one 
should look for linear conditions that can be solved explicitly. If 
one is allowed to make a gauge choice $(\d_1X\cdot\d_2X)(\s,0)=0$, it 
combines with eq. (\FirstQuasiNeumann) to $\d_2X^{a'}(\s,0)=0$, a 
linear Neumann condition. This is in fact a valid gauge choice 
corresponding to the generator $(\d_2\T_1)(\s,0)$. This can be seen by 
examining which diffeomorphisms preserve the boundary conditions. 
The coordinates transform under a diffeomorphism with parameter $f^i$ 
as $\delta_fX=f^i\d_iX$. The presence of coordinates with Dirichlet 
boundary conditions implies that $f^2(\s,0)=0$. Expansion to linear 
order in $\r$: $f^1=f^1_{(0)}(\s)+\r f^1_{(1)}(\s)+\O(\r^2)$, 
$f^2=\r f^2_{(1)}(\s)+\O(\r^2)$ gives the transformation 
$\delta_{f}\d_2X^{a'}(\s,0)=f^1_{(1)}(\s)\d_1X^{a'}(\s,0)$, showing that 
exactly this transformation may be ``used up'' to complete the Neumann 
boundary conditions. This procedure is carried out step by step, so 
that in the end all odd derivatives of $\T_1$ are gauged fixed by
$(\d_1X\cdot\d_2^{2n+1}X)(\s,0)=0$. It is also necessary for the 
non-linearities not to interfere with the time development of the 
Dirichlet boundary conditions. The resulting system has pure linear 
Neumann conditions for $X^{a'}$, $P^{a'}$ (all odd derivatives with 
respect to $\r$ vanish) and Dirichlet conditions for 
$X^{a''}$, $P^{a''}$ (even derivatives vanish). The remaining space 
diffeomorphisms consist of even $f^1$ and odd $f^2$.

Specialising to a membrane with the topology of a finite cylinder 
(some comments on topology will be made in the last section), 
$0\leq\r\leq\pi$, and Dirichlet conditions at the two boundaries
$X^{a''}(\s,0)=0$, $X^{a''}(\s,\pi)={L^{a''}\over2}$,
the coordinates can be expanded in modes:
$$
\eqalign{
X^{a'}(\s,\r)&=x^{a'}+\fr\pi\!\!\!\!
\sum_{n=-\infty}^\infty\sum_{m=1}^\infty x_n^{a'}e^{in\s}\cos m\r\komma\cr
P^{a'}(\s,\r)&=\fr{2\pi^2}p^{a'}+\fr\pi\!\!\!
\sum_{n=-\infty}^\infty\sum_{m=1}^\infty p_n^{a'}e^{in\s}\cos m\r\komma\cr
}\Eqn\OModeExpansionA
$$
$$
\eqalign{
X^{a''}(\s,\r)&=\Fr{L^{a''}}{2\pi}\!\r+\fr\pi\!\!\!\!
\sum_{n=-\infty}^\infty\sum_{m=1}^\infty x_n^{a''}e^{in\s}\sin m\r\komma\cr
P^{a''}(\s,\r)&=\fr\pi\!\!\!
\sum_{n=-\infty}^\infty\sum_{m=1}^\infty p_n^{a''}e^{in\s}\sin m\r\punkt\cr
}\Eqn\OModeExpansionB
$$
There is still a large gauge invariance, the algebra of 
diffeomorphisms preserving this class of functions. 

For future reference, we also give the conventions for the mode 
expansion of a winding membrane,
$$
\eqalign{
X^{a}(\s,\r)&=\Fr{L^{a}}{2\pi}\!\r+\fr{2\pi}\!\!\!
\sum_{m,n=-\infty}^\infty x_n^{a}e^{in\s+im\r}\komma\cr
P^{a}(\s,\r)&=\fr{2\pi}\!\!\!
\sum_{m,n=-\infty}^\infty p_n^{a}e^{in\s+im\r}\punkt\cr
}\Eqn\WModeExpansion
$$

\section\Truncation{Consistent truncation and affine algebras}When 
considering 
the classical or quantum dynamics of an open membrane, one may aim at 
different kinds of descriptions. Two directions that immediately come into 
mind are: 

\itemitem{{\it i})}{Trying to get rid of the coordinates obeying 
Dirichlet conditions and in that way obtain something similar to a 
kind of string theory (although not conformally invariant). If there 
is only one Dirichlet coordinate, as in the Ho\v rava--Witten 
construction [\HoravaWitten], the gauge symmetry suffices to 
eliminate it completely.}

\itemitem{{\it ii})}{Working with the kinds of structures 
that are inherent to membrane 
theory, \ie, choose light-cone gauge and try to truncate the 
residual gauge symmetry, the algebra of area-preserving 
diffeomorphisms [\BSTT], and the coordinates to some kind 
of matrix theory.}

\noindent The aim of this section is to demonstrate how a consistent
truncation to an affine algebra manages to combine these approaches.

Light-cone gauge amounts 
to choosing $P^+=\fr{4\pi^2}p^+$ (closed toroidal membrane) or 
$P^+=\fr{2\pi^2}p^+$ (open cylindrical membrane) 
for the hamiltonian constraint 
and $\d_iX^+=0$ for $\T_i$. Since the gauge choice for the space part 
is a total derivative, there is the remaining gauge symmetry 
$\e^{ij}\d_i\T_j=\{P_I,X^I\}\approx0$ (here, $a=(+,-,I)$), generating 
area-preserving diffeomorphisms.

The algebra of area-preserving diffeomorphisms is 
generated by functions $f(\xi)$
on the space-sheet with the bracket 
(\PoissonBracket), together with global vector fields with 
holonomies around A- and B-cycles with the obvious action on each 
other and on the functions. Consider first the algebra of function 
under the Poisson bracket. It is a well known fact that this algebra, 
in a certain sense, is $su(\infty)$ [\deWitHoppeNicolai,\FairlieZachos]. 
The reason is 
clear from the truncation below. To get to matrix theory, 
with a finite-dimensional gauge algebra $su(N)$, there is a 
well-defined mathematical procedure. One introduces a 
non-commutativity $[\s,\r]=i\theta$ on the membrane encoded in the 
(Weyl-ordered) star product
$$
f\star g=
fe^{{i\over2}\theta\e^{ij}\buildrel{\ss\leftarrow}\over\d\!\!{}_i
\buildrel{\ss\rightarrow}\over\d\!\!{}_j}g\punkt\eqn
$$
The Fourier modes of definite momentum then commute as
$$
[e^{ik_i\xi^i},e^{ik'_i\xi^i}]_\star=-2i\sin(\half\theta k\times k')
e^{i(k_i+k'_i)\xi^i}
\komma\Eqn\StructureConstants
$$
where $k\times k'=\e^{ij}k_ik'_j$.
The space-sheet Poisson bracket is recovered as the linear term in the limit 
$\theta\rightarrow0$,
$$
\{f,g\}=-i\lim_{\theta\rightarrow0}\theta^{-1}[f,g]_\star\punkt
\Eqn\PBfromComm
$$
If $\theta$ is ``rational'', $\theta=\Fr{2\pi q}N$ where $q$ and $N$ 
are coprime integers, the sine function
in the structure constants will have zeroes. The functions
$e^{isN\s}$ and $e^{irN\r}$ ($s,r\in\Z$) commute with all other 
functions, they are central elements in the algebra. This means that 
they can be consistently modded out from the algebra of function under 
the star product, since left and right multiplication coincide on all 
functions. If the star products by $e^{iN\s}$ and $e^{iN\r}$ are 
identified with the identity operator, one obtains the equivalences
$$
\eqalign{
e^{i(k+N)\s+il\r}&\approx (-1)^le^{ik\s+il\r}\komma\cr
e^{ik\s+i(l+N)\r}&\approx (-1)^ke^{ik\s+il\r}\punkt\cr
}\eqn
$$
The star-commutator algebra after this ``consistent truncation'' is 
$su(N)$. We will restrict to odd values $N=2M+1$, when the square of 
momenta can be placed symmetrically around the origin, $|k|,|l|\leq 
M$, and to $q=1$ (all $q$'s not dividing $N$ are equivalent).
Effectively, non-commutativity produces finite translations from 
multiplication with exponentials, and if a finite number ($N$) 
of these add up to an integer number of periods there is something 
very similar to a discretisation of the circle. Of course, if a 
coordinate is both discrete and periodic, so is the momentum.

If one restricts to functions that vanish at $\r=0,\pi$ (sine 
functions) occurring in the expansion of a coordinate with Dirichlet 
boundary conditions, the algebra is an orthogonal algebra 
$so(2M+1)=B_M$ [\KimRey].

Let us now come to the point of this section. In a membrane theory, 
the functions we manipulate and truncate are the actual coordinates.
If some of them have an expansion as in eqs. (\OModeExpansionB) and
(\WModeExpansion) with 
non-zero $L$'s, \ie, if the membrane is winding or stretched between two 
hypersurfaces (branes) that are not on top of each other, the linear 
function representing the winding or stretching does not extend to a 
single-valued function on the circle $\r\in[0,2\pi]$. 
We will work for a little while with winding membranes instead of 
stretched, just because the exponentials are easier to deal with, and 
then turn back to the sine and cosine functions in the expansion in eqs. 
(\OModeExpansionA), (\OModeExpansionB).
The stretching/winding
introduces a new element $\rr$ in the algebra. Its star-adjoint action 
on functions is identical to the action of the vector field 
$-i{\theta\over2\pi}\d_1$, 
namely, $[\rr,e^{ik\s}]_\star={k\theta\over2\pi}e^{ik\s}={k\over N}e^{ik\s}$. 
We conclude that it 
is no longer consistent to identify functions modulo $e^{iN\s}$, since 
the functions we previously considered equivalent have different 
eigenvalues for $\rr$ \foot\dagger{This fact was noted in 
ref. [\deWitPeetersPlefka].}. The only central elements remaining 
are $e^{irN\r}$.
Even if the functions $e^{isN\s}$ are no longer central, they form an
abelian ideal of the star commutator algebra, and one can restrict to the
quotient by this ideal, which amounts to simply discarding these terms in
the mode expansion.
We have an effective ``discretisation'' in the $\r$ direction, while $\s$ 
remains ``continuous''.

The linear function $\rr$ acts as an outer derivation on the periodic 
functions. This follows directly from the fact that it acts with the 
same star commutator and that it is not periodic, although the result 
is. As an outer derivation, 
it can be used to construct a non-trivial central extension:
$$
[f,g]'=[f,g]_\star+\gamma\,\tr(f\,[\,\rr,g]_\star)\punkt\eqn
$$
Clearly, in any consistent truncation containing $\rr$ it will 
remain an outer derivation. Since all derivations of 
finite-dimensional Lie algebras are inner, and they have no extensions 
of non-trivial cohomology, this shows that one can not 
hope to reach a matrix theory with finite-dimensional matrices.

The algebra resulting after the identification 
$$
e^{ik\s+i(l+N)\r}\approx (-1)^ke^{ik\s+il\r}\Eqn\Identification
$$ 
and exclusion of the functions $e^{isN\s}$
is infinite-dimensional.
It has finite rank $N=2M+1$, one set of Cartan generators can be taken as 
$e^{\pm ik\r}$, $k=1,\ldots M$, and $\rr$. This algebra is an affine 
$su(N)$, or $A_{N-1}^{(1)}$. What is a little confusing about the commutators 
(\StructureConstants) is that it is hard to see how the $su(N)$ 
subalgebra is embedded. To find this embedding is essentially the 
same problem as to construct a relation between $\rr$ and the standard 
derivation of the affine algebra. Modulo normalisation, they 
can only differ by an inner derivation. 

The structure becomes clear in an example. Let us consider $N=3$, with
Cartan generators chosen to $e^{\pm i\r}$ and $\rr$. In order to find the 
root system, we diagonalise $h=e^{i\r}$. The real components of this 
generator are orthogonal, so the result will be immediately possible 
to visualise in a euclidean plane. 
$h$ acts on the three functions 
$\{e^{in\s+im\r}\}_{m=-1,0,1}$ for fixed $n$ as the matrix
$$
h^{(n)}=2i\sin{n\pi\over 3}
\left[\matrix{0&0&(-1)^n\cr
              1&0&0\cr
	      0&1&0\cr}\right]\punkt\eqn
$$
The matrix in brackets is easily diagonalisable, and gives the 
eigenvalues $\{\mu\}=\{-1,e^{\pm{\pi i\over3}}\}$ for $n$ odd and minus the 
same numbers for $n$ even. When this is multiplied by the prefactor 
(which in this case is a sign factor up to normalisation), the result 
is that up to normalisation of $h$ the eigenvalues are zero for 
$n=0\mod3$, $\{\mu\}$ for $n=1\mod3$ and $\{-\mu\}$ for $n=2\mod3$.
With suitable normalisation of $h$ and the derivation, whose 
eigenvalues are the $n$ mode numbers, we get the root space of 
figure {\old1}. 
As simple roots we can take $\a_1$, $\a_2$ and $\a_3$. The vertical 
direction is a null direction, and the Cartan matrix of the algebra is
$$
A=\left[\matrix{2&-1&-1\cr
                -1&2&-1\cr
		-1&-1&2\cr}\right]\komma\eqn
$$
which is the Cartan matrix of affine $su(3)$, or $A_2^{(1)}$. 
The eigenvalue $n$ of the 
derivation is not the ``standard'' one used for affine algebras, so 
the embedding of the $su(3)$ subalgebra is ``tilted'', as shown in 
the figure (the whole picture may also be rotated an angle 
$\pm{2\pi\over3}$ around the vertical).
The general picture is that the $N$ simple roots of $A_{N-1}^{(1)}$ are 
found at eigenvalue ${1\over N}$ of $\rr$, and are oriented as the $N-1$
simple roots of $su(N)$ and minus the highest root. 
Since the simple $su(N)$ roots ($\a_1$ and $\a_2$ for the $su(3)$ case) 
have eigenvalue ${1\over N}$ under
the derivation $\rr$, a root at height $\ell$ 
(a sum of $\ell$ simple roots) has eigenvalue ${\ell\over N}$. The relation 
between the derivation used so far and the standard derivation of an 
affine algebra is then 
$$
d=\rr-{1\over 
N}H_\rho\komma\Fqn\StandardDerivation\StandardDerivationFoot
$$
where $H_\rho$ is the element in the $su(N)$ Cartan subalgebra whose 
eigenvalues are the heights. This is achieved for the Weyl vector
$\rho=\sum_{i=1}^{N-1}\lambda_{(i)}$, the sum of all fundamental 
weights of $su(N)$. In general (see \eg\ ref. [\FuchsSchweigert]), 
the the derivation for an affine algebra based on a finite-dimensional
Lie algebra ${\frak g}$
introduces a gradation of ${\frak g}$ into eigenspaces under some 
automorphism. If this automorphism is inner, \ie, generated by an (inner)
derivation, the algebra is isomorphic, via a redefinition of the derivation
as in eq. (\StandardDerivation), to an untwisted affine algebra. Twisted
affine algebras are classified by outer automorphisms of ${\frak g}$.

So far, we have strictly speaking been talking about winding closed
membranes, not stretched open membranes. The boundary conditions, as 
shown in section \BoundaryConditions, enforce the transverse, Dirichlet, 
coordinates $X^{a''}$ to contain the stretchings $L$ and sine functions, 
while the longitudinal, Neumann, ones $X^{a'}$ are expanded in cosines
(the light-cone $+-$ directions eliminated in the passage to light-cone gauge
are of course in the longitudinal sector).
Exactly as the split into sines and cosines breaks $su(N)$ to $so(N)$,
where the sines span the adjoint representation and the cosines
the symmetric traceless representation, the same happens for the affine
algebra (the tracelessness corresponds to the removal of the ideal 
$e^{isN\s}$).

\section\OpenWindingMembranes{Affine matrix theory for winding or stretched
membranes}Let us forget the translational modes for a while, 
they commute with all
elements of the algebra, and any (localised) 
quantum mechanical state we get will in
the end be promoted to a field (in the light-cone gauge) 
depending on $x^{I'}$.
After the consistent truncation has been performed, 
we are in a position to write down a mode expansion for the coordinates
and momenta of a winding or stretched membrane. In the winding case, the
gauge algebra is $A_{N-1}^{(1)}$ with generators $\{T^\A_n,d\}$, where $\A$ is
an adjoint $su(N)$ index and $n\in\Z$ are the eigenvalues of $d$. We take 
the generators to be normalised as 
$\Tr(T^\A_m T^\B_n)=\tr(T^\A T^\B)\delta_{m+n,0}=\delta^{\A\B}\delta_{m+n,0}$, 
corresponding to the orthonormality of the trigonometric functions.
The coordinates and momenta, truncated from eq. (\WModeExpansion), are
$$
\eqalign{
X^I&=L^Id+\sum_{n=-\infty}^\infty x^{I\A}_nT^\A_n\komma\cr
P^I&=\sum_{n=-\infty}^\infty p^{I\A}_nT^\A_n\komma\cr
}\eqn
$$
with Poisson brackets
$\{x^{I\A }_m,p^{J\B}_n\}_{PB}=\delta^{IJ}\delta^{\A\B}\delta_{m+n,0}$.
Equivalently, we can think of $n$ as a mode number of the expansion in a
coordinate $\s$ \foot\star{Notice that this $\ss\s$ 
is {\xit not} the same $\ss \s$ as
the original membrane coordinate, since we performed a ``de-twisting'' of
the algebra in eq. (\StandardDerivationFoot). 
While the spectrum of $\ss\rr$ is fractional
with spacing $\sss{1\over N}$, $\ss d$ has integer spectrum. 
In this way, the new $\ss \s$
describes a circle with $\ss N$ times smaller circumference 
than the old $\ss\s$,
and, depending on the height $\ss\ell$, different $\ss su(N)$ generators are 
shifted by different fractions $\ss{\ell\over N}$ to become integer modes in
the new $\ss\s$.},
$$
\eqalign{
X^I(\s)&=-iL^I{\d\over\d\s}+\fr{\sqrt{2\pi}}\!\!\sum_{n=-\infty}^\infty 
	x^{I\A}_nT^\A e^{in\s}\komma\cr
P^I(\s)&=\fr{\sqrt{2\pi}}\!\!\sum_{n=-\infty}^\infty 
	p^{I\A}_nT^\A e^{in\s}\komma\cr
}\eqn
$$
Here, we have treated winding and non-winding coordinates together.
In the case of a stretched membrane, as mentioned, the Dirichlet coordinates
go into the adjoint representation of the affine algebra $B_M^{(1)}$ and
the Neumann ones into the symmetric traceless representation.
We thus have the expansion
$$
\eqalign{
X^{I'}(\s)&=\fr{\sqrt{2\pi}}\!\!\sum_{n=-\infty}^\infty 
	x^{I'{AB}}_nS^{AB}e^{in\s}\komma\cr
P^{I'}(\s)&=\fr{\sqrt{2\pi}}\!\!\sum_{n=-\infty}^\infty 
	p^{I'{AB}}_nS^{AB}e^{in\s}\komma\cr
}\eqn
$$
$$
\eqalign{
X^{I''}(\s)&=-iL^{I''}{\d\over\d\s}
	+\fr{\sqrt{2\pi}}\!\!\sum_{n=-\infty}^\infty 
	x^{I''{AB}}_nT^{AB}e^{in\s}\komma\cr
P^{I''}(\s)&=\fr{\sqrt{2\pi}}\!\!\sum_{n=-\infty}^\infty 
	p^{I''{AB}}_nT^{AB}e^{in\s}\komma\cr
}\eqn
$$
where $A,B$ are vector indices of $so(N)$, $T^{AB}$ is a normalised basis for
the adjoint representation and $S^{AB}$ for the traceless symmetric
representation.

According to the correspondence between the space-sheet Poisson 
bracket and the star commutator (eq. (\PBfromComm) without the limit), 
the latter now being an ordinary commutator,
$\{f,g\}\rightarrow-{iN\over2\pi}[f,g]$. 
Since the constant modes have been left out of the discussion, the 
light-cone hamiltonian gives the mass operator for a winding membrane:
$$
\fr2M^2=(2\pi)^2
\Tr\Bigl\{\fr2P^IP^I+\fr4T^2{N^2\over(2\pi)^2}[X^I,X^J][X^I,X^J]\Bigr\}
\komma\eqn
$$
where the trace is over the affine algebra. Choose a basis where one 
of the coordinates, $Y$, has a winding $L$ and the others, $X^R$, 
are periodic. Rescaling to dimensionless variables, 
$X^I\rightarrow LX^I$, $P^I\rightarrow L^{-1}P^I$,
identifies Y with 
$-i$ times a covariant $\s$-derivative, $Y=-iD=-i\d_\s+A$. 
We denote the conjugate momentum $E$. The affine trace is expressed as
$\Tr=\int d\s\tr$, so that
$$
\eqalign{
\fr2M^2={(2\pi)^2\over L^2}\int d\s
\tr\Bigl\{\fr2&\bigl(E^2+P^RP^R\bigr)\cr
+\fr2{T^2L^6}{N^2\over(2\pi)^2}
&\bigl(-DX^RDX^R+\fr2[X^R,X^S][X^R,X^S]\bigr)\Bigr\}\punkt\cr
}\eqn
$$
If we define the dimensionless coupling constant 
$g=\sqrt{N\over2\pi TL^3}$,
$$
\eqalign{
\fr2\left({gLM\over2\pi}\right)^2
=\int d\s\tr\Bigl\{\fr2g^2&\bigl(E^2+P^RP^R\bigr)\cr
+\fr2g^{-2}&\bigl(-DX^RDX^R+\fr2[X^R,X^S][X^R,X^S]\bigr)\Bigr\}\punkt\cr
}\eqn
$$
This is the hamiltonian of $d=2$ Yang--Mills with coupling constant 
$g$ coupled to $D-3$ adjoint scalars 
with a quartic potential in the gauge $A_0=0$, \ie, of $d=D-1$ 
Yang--Mills dimensionally reduced to $d=2$. 
The consistent truncation performed has thus led to matrix string theory 
[\DVV].

When the membrane is stretched, the truncated algebra is affine $so(N)$.
The Dirichlet coordinates transform in the adjoint representation and 
the Neumann ones in the traceless symmetric representation. 
We get the expression for the mass operator:
$$
\eqalign{
\left({gLM\over2\pi}\right)^2
=\int d\s\tr\Bigl\{\fr2g^2&\bigl(E^2+P^RP^R+P^{I'}P^{I'}\bigr)\cr
+\fr2g^{-2}&\bigl(-DX^RDX^R-DX^{I'}DX^{I'}
+\fr2[X^{I'},X^{J'}][X^{I'},X^{J'}]\cr&+\fr2[X^R,X^S][X^R,X^S]
+[X^R,X^{I'}][X^R,X^{I'}]\bigr)\Bigr\}\punkt\cr
}\eqn
$$
This is the hamiltonian of $d=2$ Yang--Mills with gauge group $so(N)$ 
coupled to $D-p-2$ adjoint scalars and $p-1$ scalars in the 
symmetric traceless representation
\ie, the dimensional reduction to $d=2$ of $d=D-p$ Yang--Mills 
coupled to to $p-1$ symmetric traceless scalars.

\section\SuperMembranes{Supermembranes}In order to treat 
supersymmetric membranes, the boundary conditions for the fermions 
have to be analysed. This has been done in detail in 
ref. [\EzawaMatsuoMurakami]. There it was shown that, under the assumption
that some supersymmetry remains unbroken, the possible values of $p$ are
{\old1}, {\old5} and {\old9}. 
It is not clear whether $p=1$ has some natural interpretation in 
M-theory; the other two cases correspond to membranes ending on M{\old5}-branes
or on end-of-the-world {\old9}-branes. 

The case of {\old9}-branes is slightly complicated. It is straightforward
to verify that when the light-cone rotation algebra $so(9)$
is broken by the stretching to $so(8)$, the spinor 
$16\rightarrow8_s\oplus8_c$, of which one spinor is odd over the boundary
and the other one even, so they go into the adjoint and the symmetric
traceless representations of the affine algebra. 
It was argued in ref. [\KimRey] that this
boson-fermion mismatch should be compensated by a twisted sector in the
fundamental representation. It is also known that the classical anomaly
cancellation of the supermembrane demands degrees of freedom residing at the 
boundary, corresponding to an $E_8$ WZW model 
[\CederwallOpenMembrane,\BraxMourad].
We hope to be able to come back to this point,
and turn to supermembranes ending on M{\old5}-branes.

The light-cone rotation group of the supermembrane, $so(9)$, is then 
broken into 
$so(4)_{||}\oplus so(4)_\perp\cong su(2)^4$, one of the $so(4)$'s
being the little group of the longitudinal light-cone and the other one
rotations between the non-stretched transverse directions.
The light-cone spinor decomposes as 
$16\rightarrow(1,2;1,2)\oplus(1,2;2,1)\oplus(2,1;1,2)\oplus(2,1;2,1)$, 
The boundary conditions, as derived in ref. [\EzawaMatsuoMurakami],
force these spinors  
to be odd, odd, even and even, respectively, over the boundary.
The odd ones combines with the gauge field (the stretching coordinate) and
the the $D-p-2=4$ odd longitudinal scalars in $(2,2,1,1)$ 
into one supermultiplet and the even ones
with the $p-1=4$ even transverse unstretched scalars in $(1,1,2,2)$ 
into another multiplet.
The global $su(2)^4$ symmetry plays the r\^ole of an R-symmetry of the model.
The mass operator is such that $(gLM/2\pi)^2$ is the
hamiltonian of ${\cal N}=(1,0)$, 
$d=6$ super-Yang--Mills theory with gauge group
$so(N)$, coupled to a hypermultiplet in the symmetric 
traceless representation, dimensionally reduced to $d=2$, with 
coupling constant $g$ (recall that adding hypermultiplets to $d=6$ 
super-Yang--Mills does not introduce an additional coupling constant).

\section\conclusions{Discussion}We have shown how, in the presence of
winding or stretching in one direction, a membrane can be truncated to
an affine matrix theory, a matrix string theory.
There are obvious applications in any situation where winding or stretched
membranes occur. M-theory compactified on $S^1$, type \II A string theory,
is one such example that has already been investigated [\DVV].
Other M-theory backgrounds that should be examined are 
the one in the Ho\v rava--Witten construction and backgrounds with 
parallel M{\old5}-branes. 

The modern interpretation of supermembrane theory implies that the 
topology of the membrane is irrelevant. Classically, a membrane can
grow ``spikes'' at infinitesimally low energy cost. In the matrix
truncation this is reflected in the existence of flat directions in
the quartic potential. This flatness is not lifted by quantum effects
in the supersymmetric membrane. Already a first-quantised supermembrane
contains ``multi-membrane'' states. The properties are nicely reflected
in the truncation to matrices of finite size, where all topological
information about the membrane is lost, which should presumably be
seen as physically correct.
One may ask the question whether or not this is true for membranes with
boundaries. The question actually disintegrates in two: does stretching
stabilise the bulk of the membrane against topology change, and is
the topology of the boundary fixed? 

In ref. [\deWitPeetersPlefka] it was argued that a winding membrane
still has continuous spectrum, although the lowest energy is raised due
to the winding. The same should hold for a stretched membrane.
There are clearly flat directions in the potential, corresponding to
commuting elements in the affine algebra. A ground state at the bottom of
the spectrum should be BPS-saturated (it would be very interesting if the 
Witten index can be calculated). Then, on the other hand, it seems
clear that the membrane is actually locked to the boundary branes, since
it otherwise could have disintegrated into some closed membrane states.
It is tempting to conjecture that the continuity of the spectrum
stems from a description of ``multi-string'' states, as seen from the
$p+1$-dimensional world-volume, and not from interaction with bulk
degrees of freedom. 

It is possible that the model described in section \SuperMembranes\ can
be used to describe ${\cal N}=(2,0)$ theories on systems of
parallel M{\old5}-branes. The situation that has been considered here
corresponds to {\it one} off-diagonal element represented by a membrane
stretched between a fixed pair among the branes, and it is already obvious
that diagonal and non-diagonal elements behave very differently (this can
also be seen in an inherently {\old6}-dimensional approach to constructing
``non-abelian tensor multiplets'' [\Henningson]).
It would of course
be nice to get more information concerning the relevance of topology
in this and similar cases. If the above conjecture, that already a
cylindrical membrane captures all boundary topologies, is true, it would
be part of the answer. On the other hand, on should not {\it a priori}
exclude states described by membranes ending on more than two branes, unless
it can be argued that they are already in some way included.

The separation of the branes, the stretching, introduces a length scale
which can be used to define a dimensionless coupling constant $g$
as in section \OpenWindingMembranes. It may be used for a perturbation
expansion, where small coupling (large $L$) corresponds to strongly
coupled string theory in the cases where such a correspondence
exists (type \II A and heterotic).

The non-commutativity parameter, which in string theory has a physical
meaning, acts in the present context as a device to deform the algebra
to something that allows a consistent truncation. 
Whether or not it has a more direct physical interpretation is unclear. 
It is known
that matrix theory, for finite and fixed $N$, breaks Lorentz invariance
[\deWitMarquardNicolai]. 
This fact can be traced directly to the manifestly non-covariant introduction
of non-commutativity. To our knowledge it has never been shown, however,
that Lorentz invariance has to be broken if $\theta$ is allowed to transform.
It has been discussed that the non-commutativity should be lifted to some
covariant ``{\old3}-dimensional'' structure on the membrane 
world-volume, see \eg\ refs. [\Minic,\Schaar,\BermanEtAl]. 
Another question concerning the algebraic structure of the theory is
if there is some physical situation that, \eg\ via orientifolding,
would lead to twisted affine algebras.

An interesting feature, not present in matrix theory, is that the symmetry
algebra is large enough to allow for quantum mechanical anomalies.
It would be very interesting if demanding that the symmetry remains unbroken
by quantum effects could lead to a calculation of a critical dimension
for the supermembrane.

\acknowledgements
This work is supported in part by the Swedish Science Council and by
EU contract HPRN-CT-2000-00122. The author wants to thank Bengt E.W. 
Nilsson and M\aa ns Henningson for discussions.

\vfill\eject

\refout 

\vfill\eject

\centerline{\epsffile{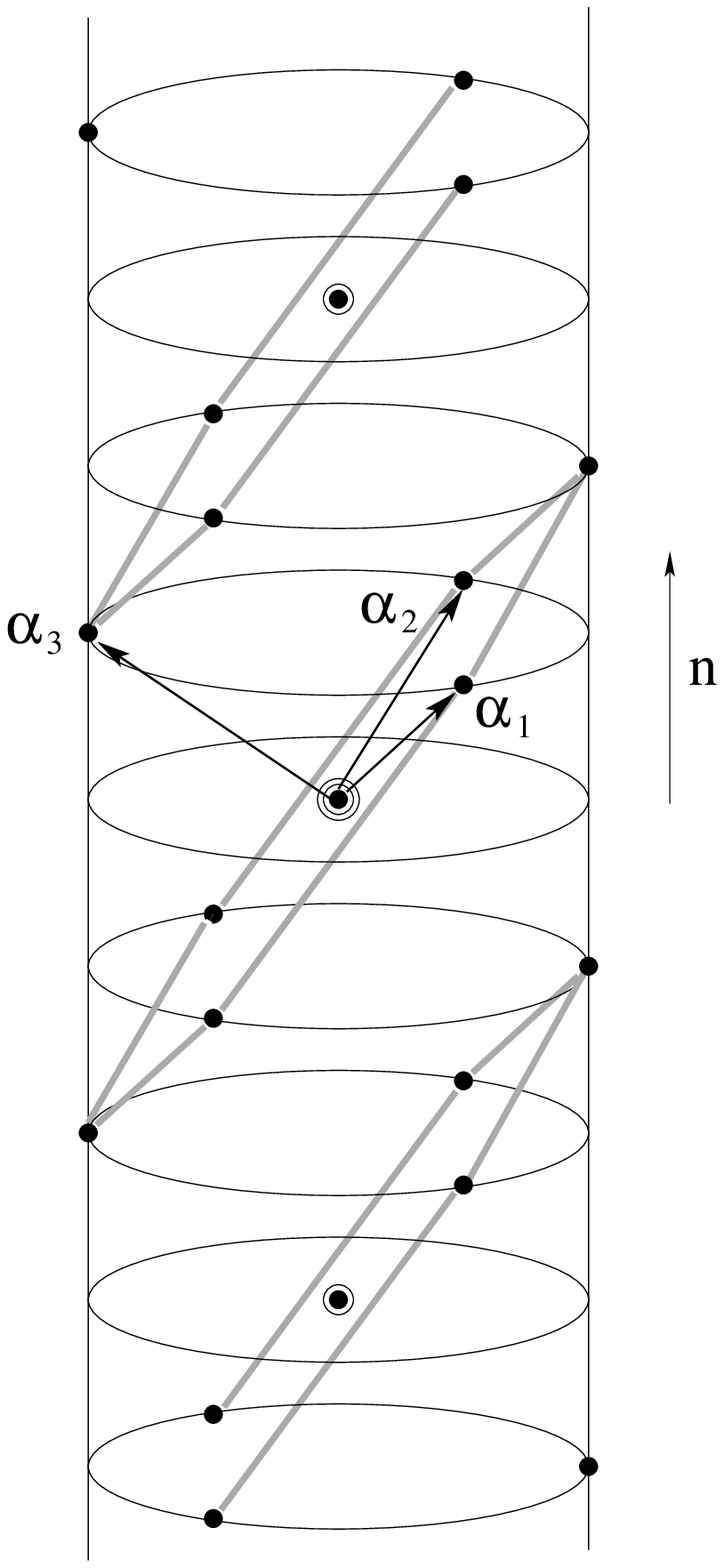}}
\vskip2\parskip
\centerline{\it Figure 1}

\end